\pdfoutput=1
\documentclass[preprint,showpacs,preprintnumbers,amsmath,amssymb,floatfix,endfloats*]{revtex4}

\usepackage{graphicx}
\usepackage{dcolumn}
\usepackage{bm}
\usepackage{amsfonts}

\DeclareMathAlphabet{\mathsfsl}{OT1}{cmr}{bx}{it}
\begin{document}
\title{The potential energy states and mechanical properties of thermally cycled binary glasses }
\author{Nikolai V. Priezjev$^{1,2}$}
\affiliation{$^{1}$Department of Mechanical and Materials
Engineering, Wright State University, Dayton, OH 45435}
\affiliation{$^{2}$National Research University Higher School of
Economics, Moscow 101000, Russia}
\date{\today}
\begin{abstract}

The influence of repeated thermal cycling on mechanical properties,
structural relaxation, and evolution of the potential energy in
binary glasses is investigated using molecular dynamics simulations.
We consider a binary mixture with strongly non-additive cross
interactions, which is annealed across the glass transition with
different cooling rates and then exposed to one thousand thermal
cycles at constant pressure. We found that during the first few
hundred transient cycles, the potential energy minima after each
cycle gradually decrease and the structural relaxation proceeds via
collective, irreversible displacements of atoms. With increasing
cycle number, the amplitudes of the volume and potential energy
oscillations are significantly reduced, and the potential energy
minima saturate to a constant value that depends on the thermal
amplitude and cooling rate. In the steady state, the glasses
thermally expand and contract but most of the atoms return to their
cages after each cycle, similar to limit cycles found in
periodically driven amorphous materials. The results of tensile
tests demonstrate that the elastic modulus and the yielding peak,
evaluated after the thermal treatment, acquire maximum values at a
particular thermal amplitude, which coincides with the minimum of
the potential energy.



\end{abstract}




\maketitle

\section{Introduction}

The design of novel strategies and optimization of thermal and
mechanical treatments of metallic glasses is important for accessing
a broad range of states with improved physical and mechanical
properties~\cite{Robert18}. It is generally accepted that an
elementary plastic event in disordered solids involves a collective
rearrangement of a small group of atoms or a shear
transformation~\cite{Argon79,Spaepen77}.  Due to lack of crystalline
order, the plastic deformation of metallic glasses typically occurs
via highly localized shear bands which lead to catastrophic
failure~\cite{Egami13}.    In recent years, a number of methods were
proposed to enhance plasticity of metallic glasses, such as addition
of chemical heterogeneities~\cite{Park06,KimChem18} or a soft second
phase~\cite{SchroersNat13}.  A less intrusive way to tune the
amorphous structure is to apply cryogenic thermal cycling that can
induce rejuvenation due to heterogeneity in the local thermal
expansion and therefore improve
plasticity~\cite{Ketov15,GreerSun16,Lu18,Kerscher18,Saida18,Ketov18}.
Using atomistic simulations, it was recently shown that internal
stresses due to thermal expansion can in principle trigger a plastic
event in sufficiently large systems~\cite{Barrat18}.

\vskip 0.05in

It was originally demonstrated using molecular dynamics (MD)
simulations that the effects of annealing and aging on atomic
glasses can be reversed by mechanical rejuvenation, which expels the
system from the deep energy minimum~\cite{Stillinger00}.  Later, it
was found that a single subyield cycle overages the glass by
decreasing its potential energy, while a shear cycle with large
strain rejuvenates the glass and increases its potential
energy~\cite{Lacks04}.   More recently, a number of studies have
explored the influence of periodic shear deformation on the energy
states, particle dynamics and mechanical properties of amorphous
materials~\cite{Priezjev13,Sastry13,Reichhardt13,Priezjev14,IdoNature15,
Priezjev16,Priezjev16a,Sastry17,Priezjev17,OHern17,Priezjev18,Priezjev18a,
Heuer18}.  In particular, it was shown by athermal quasistatic
simulations that following a number of training cycles, the glassy
systems evolve into periodic limit cycles, where particle
trajectories become exactly reversible, despite large-scale
collective displacements during each
cycle~\cite{Reichhardt13,IdoNature15}.  Most recently, the
structural relaxation in binary glasses subjected to periodic
temperature variations below the glass transition was investigated
using MD simulations~\cite{Priez18tcyc}. It was found that the
potential energy after one hundred cycles acquires a distinct
minimum as a function of the thermal amplitude, which correlates
with the largest yield stress at the same
amplitude~\cite{Priez18tcyc}. It remains unclear, however, whether
these results depend on the number of thermal cycles and details of
the thermal treatment.

\vskip 0.05in

In this paper, we carried out molecular dynamics simulations to
study the effect of thermal cycling on the evolution of the
potential energy and mechanical properties of binary glasses. The
amorphous samples were initially prepared by cooling across the
glass transition to a low temperature with different rates. It will
be shown that after a number of transient cycles, binary glasses
attain low energy states and the particle dynamics becomes nearly
reversible. We find that the minimum of the potential energy after
one thousand cycles and the maximum of the stress overshoot and
elastic modulus occur at the same thermal amplitude.

\vskip 0.05in

This paper is structured as follows. The description of molecular
dynamics simulations including parameter values, annealing
procedure, and temperature variation protocol are given in the next
section. The numerical results for the potential energy time series,
nonaffine displacements, and mechanical properties are presented in
section\,\ref{sec:Results}. The conclusions and outlook are provided
in the last section.

\section{MD simulations}
\label{sec:MD_Model}

In this study, we consider a binary (80:20) mixture model to
represent a bulk metallic glass. This model was originally
introduced by Kob and Andersen (KA)~\cite{KobAnd95} to study the
amorphous metal alloy $\text{Ni}_{80}\text{P}_{20}$~\cite{Weber85}.
In the KA model, the interaction between any two atoms of types
$\alpha,\beta=A,B$ is specified via the truncated Lennard-Jones (LJ)
potential:
\begin{equation}
V_{\alpha\beta}(r)=4\,\varepsilon_{\alpha\beta}\,\Big[\Big(\frac{\sigma_{\alpha\beta}}{r}\Big)^{12}\!-
\Big(\frac{\sigma_{\alpha\beta}}{r}\Big)^{6}\,\Big],
\label{Eq:LJ_KA}
\end{equation}
with the following values of the LJ parameters
$\varepsilon_{AA}=1.0$, $\varepsilon_{AB}=1.5$,
$\varepsilon_{BB}=0.5$, $\sigma_{AB}=0.8$, $\sigma_{BB}=0.88$, and
$m_{A}=m_{B}$~\cite{KobAnd95}.  The cutoff radius for the LJ
potential is set to $r_{c,\,\alpha\beta}=2.5\,\sigma_{\alpha\beta}$
throughout the study. For clarity, we express all physical
quantities in terms of the reduced LJ units of length, mass, energy,
and time as follows: $\sigma=\sigma_{AA}$, $m=m_{A}$,
$\varepsilon=\varepsilon_{AA}$, and
$\tau=\sigma\sqrt{m/\varepsilon}$.  The system consists of $48000$
large atoms of type $A$ and $12000$ small atoms of type $B$, and,
hence, the total number of atoms is $N_{tot}=60000$. For each
configuration of atoms, the total force on each atom from its
neighbors was computed using Eq.\,(\ref{Eq:LJ_KA}), and the atom
positions and velocities were advanced using the velocity Verlet
algorithm~\cite{Allen87} with the time step $\triangle
t_{MD}=0.005\,\tau$~\cite{Lammps}.

\vskip 0.05in

\begin{figure}[t]
\includegraphics[width=9.0cm,angle=0]{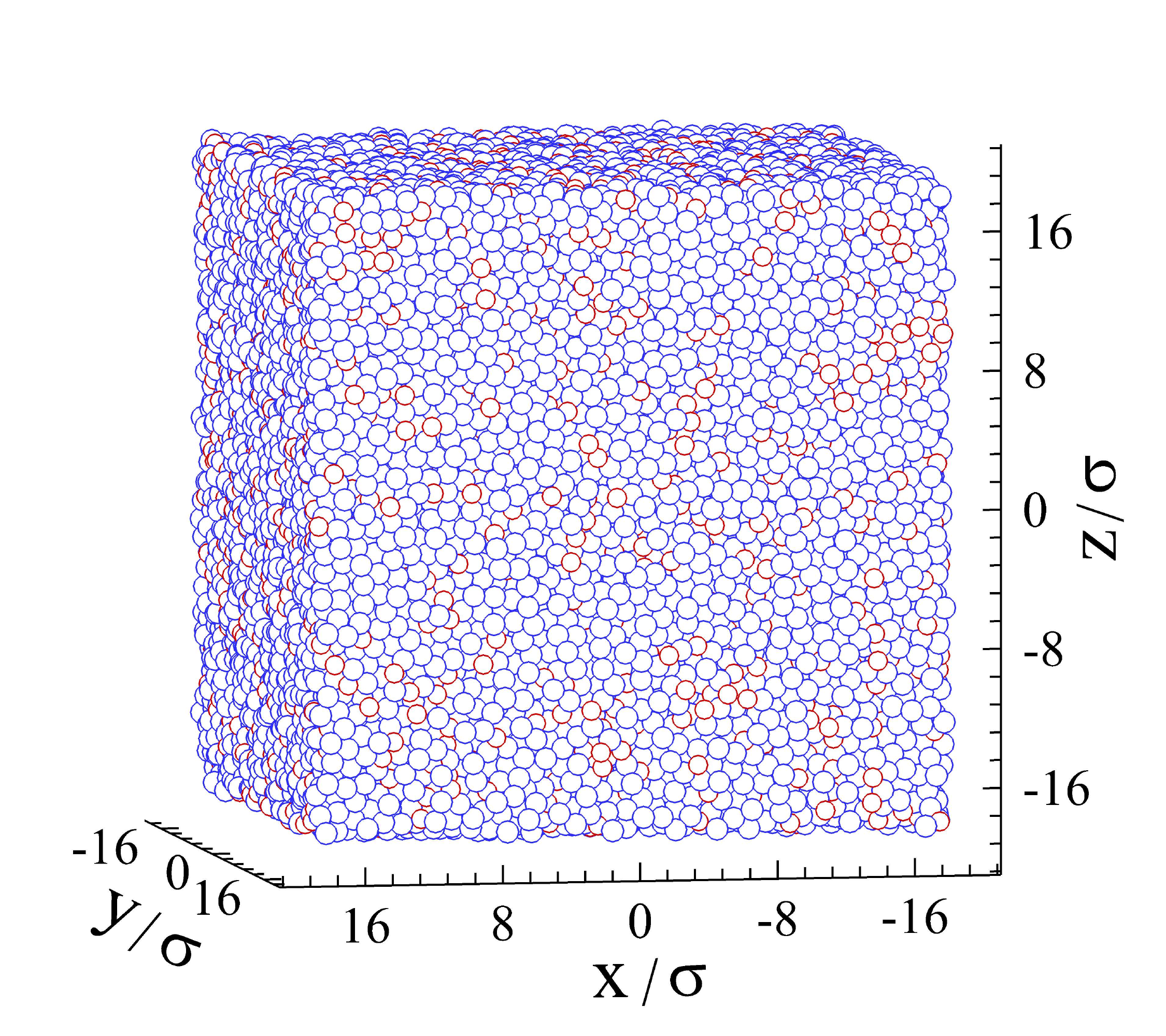}
\caption{(Color online) A snapshot of the binary Lennard-Jones glass
($60\,000$ atoms) after annealing from a liquid state to the
temperature $T_{LJ}=0.01\,\varepsilon/k_B$ with the cooling rate of
$10^{-2}\varepsilon/k_{B}\tau$.  Atoms of types $A$ and $B$ are
indicated by blue and red spheres, respectively.}
\label{fig:snapshot_system}
\end{figure}

We next briefly describe the annealing process and the temperature
variation protocols. First, the system was allowed to equilibrate in
a periodic box at zero pressure and temperature of
$0.7\,\varepsilon/k_B$. This temperature is well above the glass
transition temperature of the KA model
$T_g\approx0.435\,\varepsilon/k_B$ at the atomic density
$\rho=\rho_{A}+\rho_{B}=1.2\,\sigma^{-3}$~\cite{KobAnd95}. The
temperature, indicated by $T_{LJ}$, was maintained via the
Nos\'{e}-Hoover thermostat~\cite{Allen87}. Following the
equilibration procedure, the binary mixture was annealed into the
glass phase with the cooling rates $10^{-2}\varepsilon/k_{B}\tau$,
$10^{-3}\varepsilon/k_{B}\tau$, $10^{-4}\varepsilon/k_{B}\tau$, and
$10^{-5}\varepsilon/k_{B}\tau$. A representative snapshot of the
simulated system is shown in Fig.\,\ref{fig:snapshot_system} for the
cooling rate of $10^{-2}\varepsilon/k_{B}\tau$.  Next, the
temperature was varied piecewise linearly with the maximum values
$T_{LJ}=0.1\,\varepsilon/k_B$, $0.2\,\varepsilon/k_B$,
$0.25\,\varepsilon/k_B$, $0.3\,\varepsilon/k_B$,
$0.35\,\varepsilon/k_B$, and $0.4\,\varepsilon/k_B$ with respect to
the reference temperature of $10^{-2}\varepsilon/k_{B}\tau$. The
period of thermal cycling was set to $T=5000\,\tau$. The
post-processing analysis was performed using accumulated data for
the system dimensions, temperature, pressure components, potential
energy, and atomic configurations.

\section{Results}
\label{sec:Results}

\begin{figure}[t]
\includegraphics[width=12.0cm,angle=0]{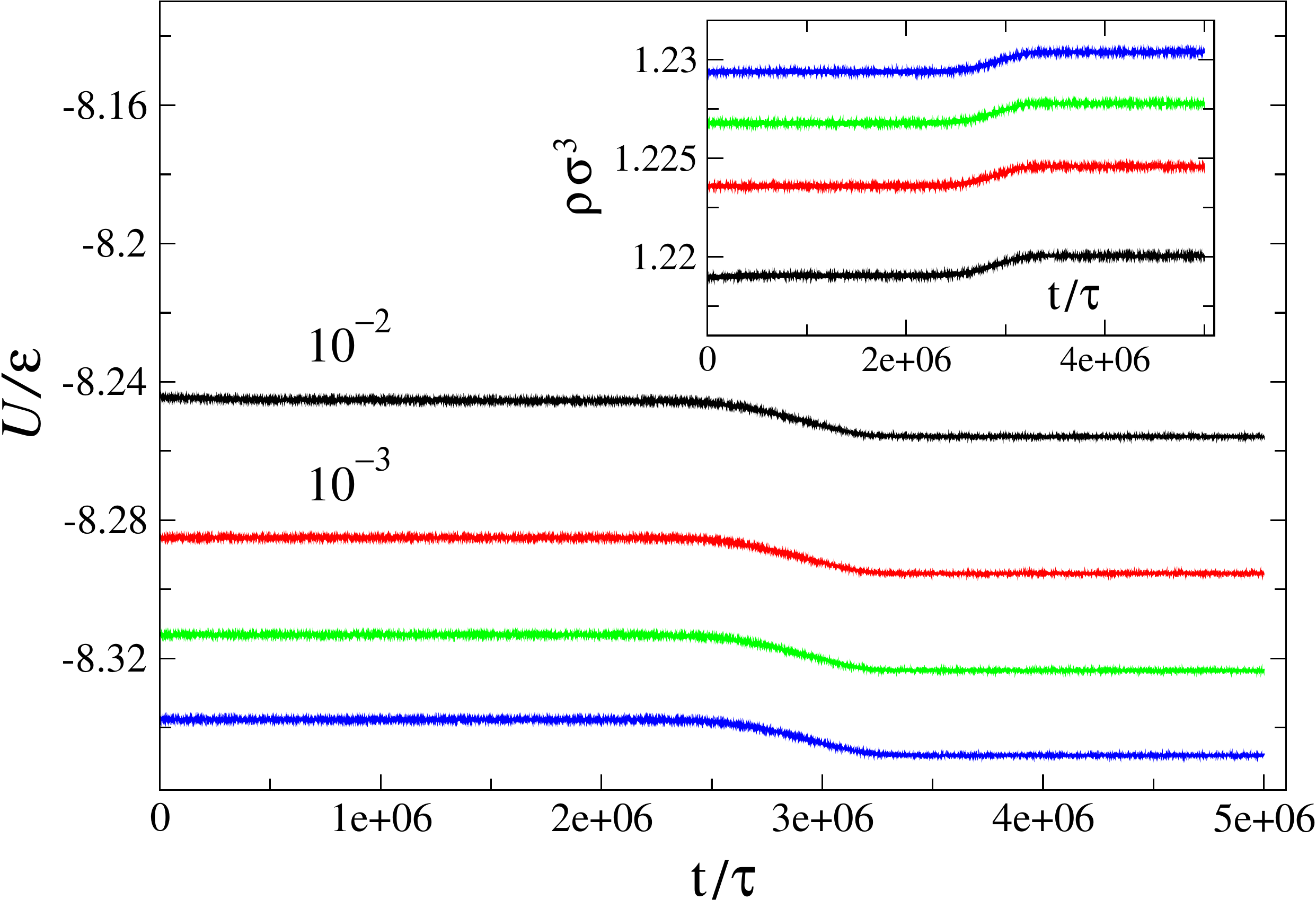}
\caption{(Color online) The time dependence of the potential energy
per atom, $U/\varepsilon$, at $T_{LJ}=0.01\,\varepsilon/k_B$ and
$P=0$ during the time interval of $5\times10^6\tau$. The glasses
were initially prepared with the cooling rates of
$10^{-2}\varepsilon/k_{B}\tau$ (black),
$10^{-3}\varepsilon/k_{B}\tau$ (red), $10^{-4}\varepsilon/k_{B}\tau$
(green), and $10^{-5}\varepsilon/k_{B}\tau$ (blue). The time
evolution of the average density for the same samples is shown in
the inset.  }
\label{fig:poten_dens_T0.01}
\end{figure}

We begin the discussion of the results by presenting in
Fig.\,\ref{fig:poten_dens_T0.01} the time dependence of the
potential energy for binary glasses annealed across the glass
transition to $T_{LJ}=0.01\,\varepsilon/k_B$ with cooling rates
$10^{-2}\varepsilon/k_{B}\tau$, $10^{-3}\varepsilon/k_{B}\tau$,
$10^{-4}\varepsilon/k_{B}\tau$, and $10^{-5}\varepsilon/k_{B}\tau$
and aged at this temperature during the time interval of
$5\times10^6\tau$. As is evident, the potential energy levels become
lower when cooling is slower, because the system has more time to
explore various minima in the potential energy landscape when
approaching the glass transition. This effect was repeatedly
observed in MD simulations of glass formers annealed below the glass
transition temperature~\cite{Ediger12}.  Somewhat surprisingly, it
can be seen in Fig.\,\ref{fig:poten_dens_T0.01} that in all cases,
the potential energy remains nearly constant during
$2.5\times10^6\tau$ and then gradually crosses over to slightly
lower levels during the time interval of about $10^6\tau$.  The
inset in Fig.\,\ref{fig:poten_dens_T0.01} shows that the decrease in
energy is accompanied with densification of glasses, since
simulations are carried out at constant pressure. The analysis of
the atomic structure, based on the pair distribution functions, did
not reveal any crystallization or enhanced short range order after
samples become denser (not shown).   We comment that most of the
long-time MD simulations of the KA model were performed either at
constant volume or near $T_g$, where the effect presented in
Fig.\,\ref{fig:poten_dens_T0.01} is absent~\cite{KobBar00}. In the
future, it might be instructive to perform a more careful analysis
of the structure during aging process at constant pressure and low
temperature.

\vskip 0.05in

\begin{figure}[t]
\includegraphics[width=12.0cm,angle=0]{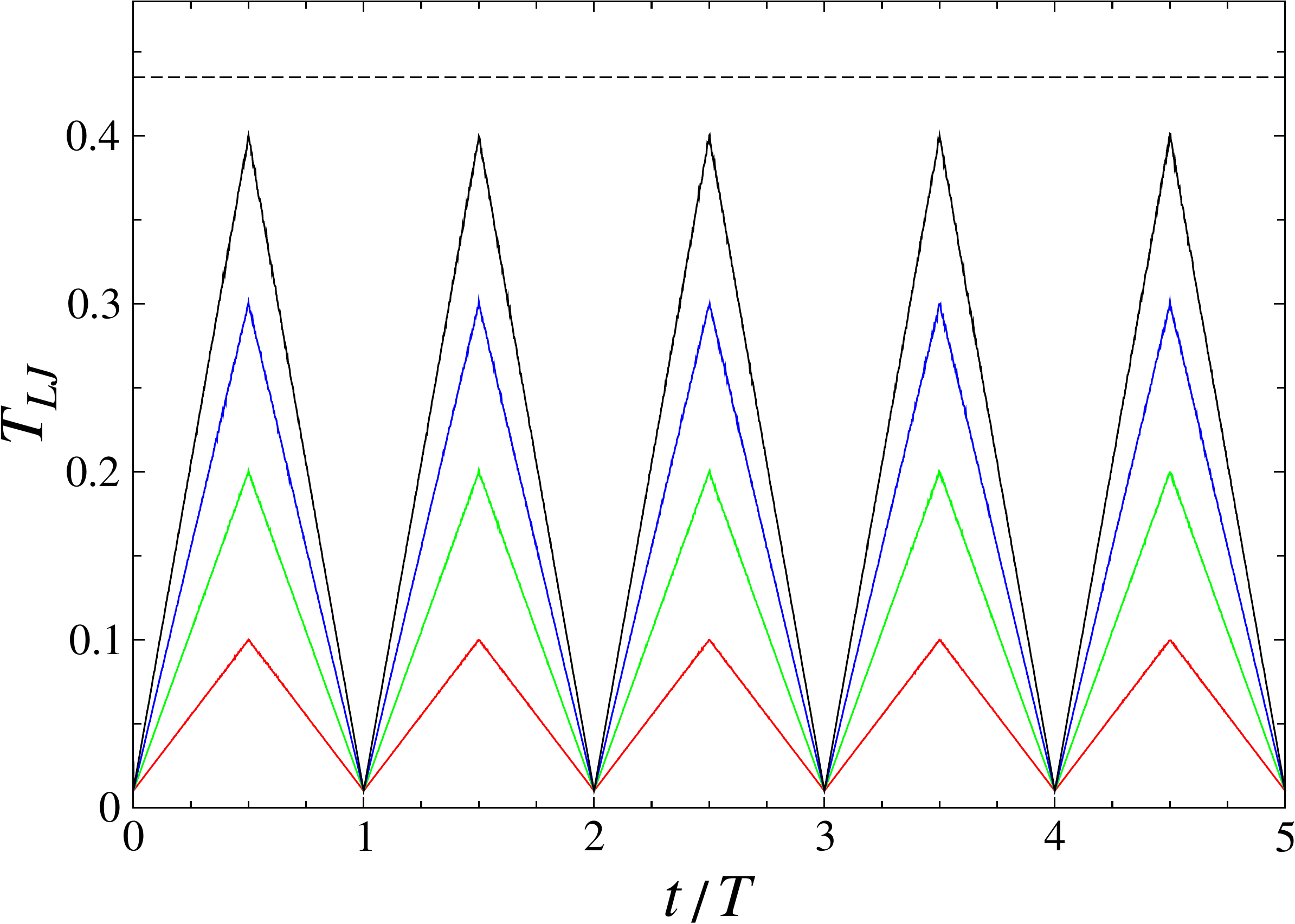}
\caption{(Color online) The temperature $T_{LJ}$ (in units of
$\varepsilon/k_B$) during the first five periods, $T=5000\,\tau$,
for the temperature amplitudes $0.1\,\varepsilon/k_B$ (red),
$0.2\,\varepsilon/k_B$ (green), $0.3\,\varepsilon/k_B$ (blue), and
$0.4\,\varepsilon/k_B$ (black). The horizontal dashed line indicates
the critical temperature of $0.435\,\varepsilon/k_B$. }
\label{fig:temp_control}
\end{figure}

As described in the previous section, the glasses were first
annealed at constant pressure to the low temperature of
$0.01\,\varepsilon/k_B$ with cooling rates
$10^{-2}\varepsilon/k_{B}\tau$, $10^{-3}\varepsilon/k_{B}\tau$,
$10^{-4}\varepsilon/k_{B}\tau$, and $10^{-5}\varepsilon/k_{B}\tau$.
The thermal treatment was then applied by repeatedly heating and
cooling the samples during 1000 cycles with the period
$T=5000\,\tau$. Examples of the temperature profiles measured during
the first five cycles are presented in Fig.\,\ref{fig:temp_control}.
During each cycle, the system gradually expands upon heating to the
maximum temperature $T_{LJ}$ and then cooled with the effective rate
of about $2\,T_{LJ}/T$. When the glass temperature become higher,
the probability of thermally activated rearrangements of atoms
increases, thus leading to structural relaxation over consecutive
cycles. In the present study, the maximum amplitude of thermal
cycling is $T_{LJ}=0.4\,\varepsilon/k_B$, just below
$T_g\approx0.435\,\varepsilon/k_B$, as it is expected that cycling
with higher thermal amplitude will result in melting and repeated
quenching to the solid phase across the glass transition.

\vskip 0.05in

\begin{figure}[t]
\includegraphics[width=12.0cm,angle=0]{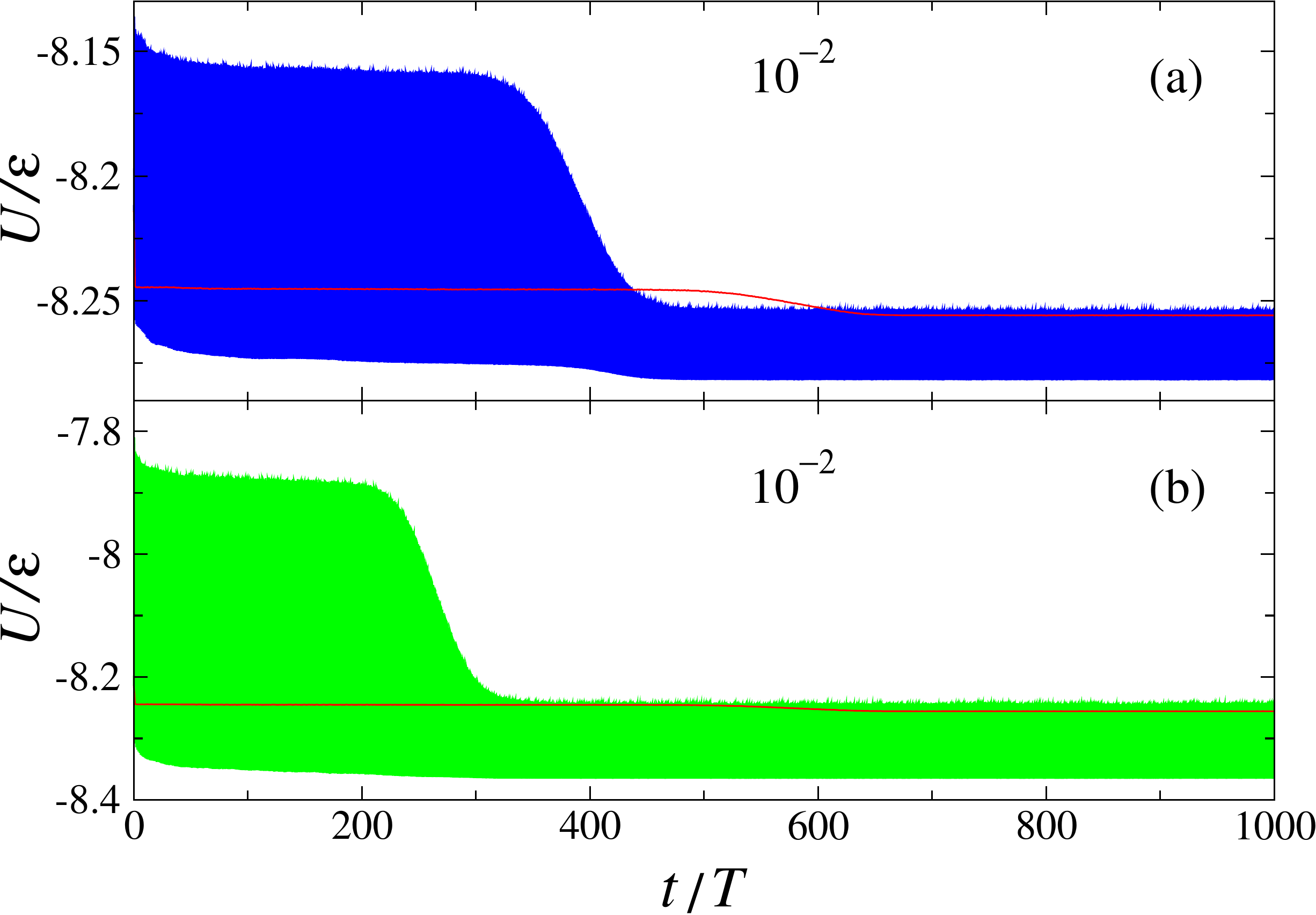}
\caption{(Color online) The potential energy during 1000 thermal
cycles with maximum temperatures (a) $0.1\,\varepsilon/k_B$ (blue)
and (b) $0.35\,\varepsilon/k_B$ (green). The potential energy of the
glass aged at $T_{LJ}=0.01\,\varepsilon/k_B$ is denoted by red
curves. The vertical scales in both panels are different. The
oscillation period is $T=5000\,\tau$. The glass was initially
prepared with the cooling rate of $10^{-2}\varepsilon/k_{B}\tau$. }
\label{fig:poten_10m2}
\end{figure}
\begin{figure}[t]
\includegraphics[width=12.0cm,angle=0]{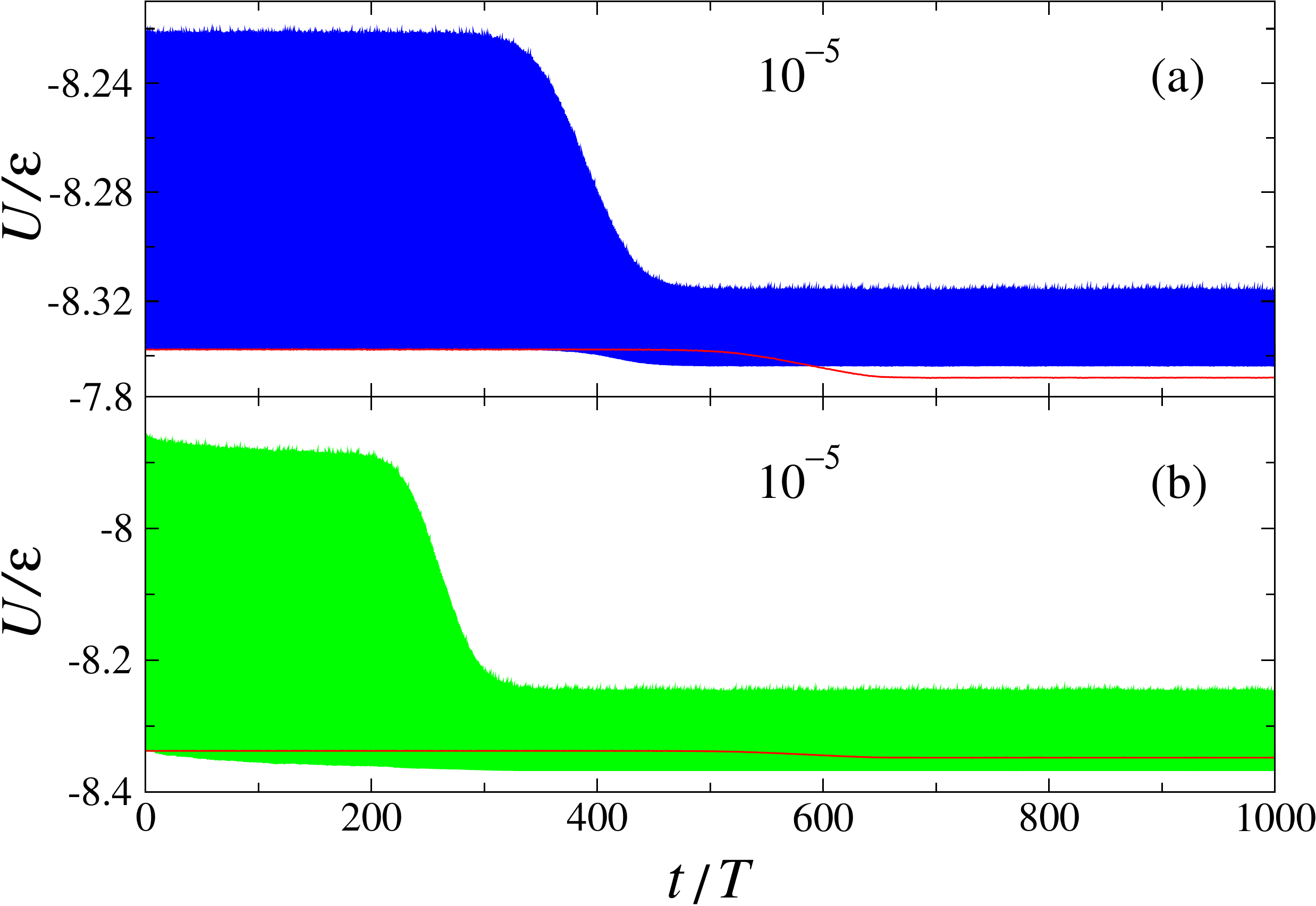}
\caption{(Color online) The potential energy series for the
thermally cycled glass with maximum temperatures (a)
$0.1\,\varepsilon/k_B$ (blue) and (b) $0.35\,\varepsilon/k_B$
(green). Before the thermal treatment, the glass was annealed with
the cooling rate of $10^{-5}\varepsilon/k_{B}\tau$. Red curves
denote the potential energy for the glass aged at
$T_{LJ}=0.01\,\varepsilon/k_B$. The period of thermal oscillations
is $T=5000\,\tau$. The vertical scales are different in both cases.}
\label{fig:poten_10m5}
\end{figure}

The representative potential energy series during 1000 thermal
cycles are shown in Fig.\,\ref{fig:poten_10m2} for poorly
($10^{-2}\varepsilon/k_{B}\tau$) and in Fig.\,\ref{fig:poten_10m5}
for well ($10^{-5}\varepsilon/k_{B}\tau$) annealed glasses and
thermal amplitudes $0.1\,\varepsilon/k_B$ and
$0.35\,\varepsilon/k_B$. For reference, the potential energies for
the aged glasses at $T_{LJ}=0.01\,\varepsilon/k_B$ are also
presented in Figs.\,\ref{fig:poten_10m2} and \ref{fig:poten_10m5} by
red curves; the same data as in Fig.\,\ref{fig:poten_dens_T0.01} for
samples annealed with cooling rates $10^{-2}\varepsilon/k_{B}\tau$
and $10^{-5}\varepsilon/k_{B}\tau$.   In can be observed in
Fig.\,\ref{fig:poten_10m2}\,(a) that the amplitude of the potential
energy oscillations is relatively large during the first 300 cycles,
and then it is gradually reduced by a factor of about three during
the next hundred cycles until a steady state in attained.  Note that
the potential energy minima after every cycle (the lower envelope of
the energy oscillations) gradually decreases during the first 500
cycles but then it remains constant in steady state.  We also
comment that similar features, as shown in
Fig.\,\ref{fig:poten_10m2}\,(a), are present for the variation of
the total volume, \textit{i.e.}, the system becomes denser and the
amplitude of the volume oscillations is reduced in the steady state
(not shown).

\vskip 0.05in


The thermal cycling with the higher maximum temperature
$0.35\,\varepsilon/k_B$, shown in Fig.\,\ref{fig:poten_10m2}\,(b),
results in qualitatively similar response; however, the transition
to the steady state occurs after about 200 cycles. Note that in both
cases presented in Fig.\,\ref{fig:poten_10m2}, the onset of the
potential energy decrease in thermally cycled glasses appears sooner
than for samples aged at the low temperature
$0.01\,\varepsilon/k_B$. The same conclusions can be deduced from
the potential energy series for the well annealed glass (cooling
rate $10^{-5}\varepsilon/k_{B}\tau$) shown in
Fig.\,\ref{fig:poten_10m5}, except that the energy levels for the
aged glass (red curves) remain closer to the energy minima of
thermally cycled glasses.   We comment that the characteristic decay
of the amplitude of potential energy oscillations after several
hundred cycles was observed in all samples and all thermal
amplitudes considered in the present study; however they are omitted
here for brevity.

\vskip 0.05in

\begin{figure}[t]
\includegraphics[width=12.0cm,angle=0]{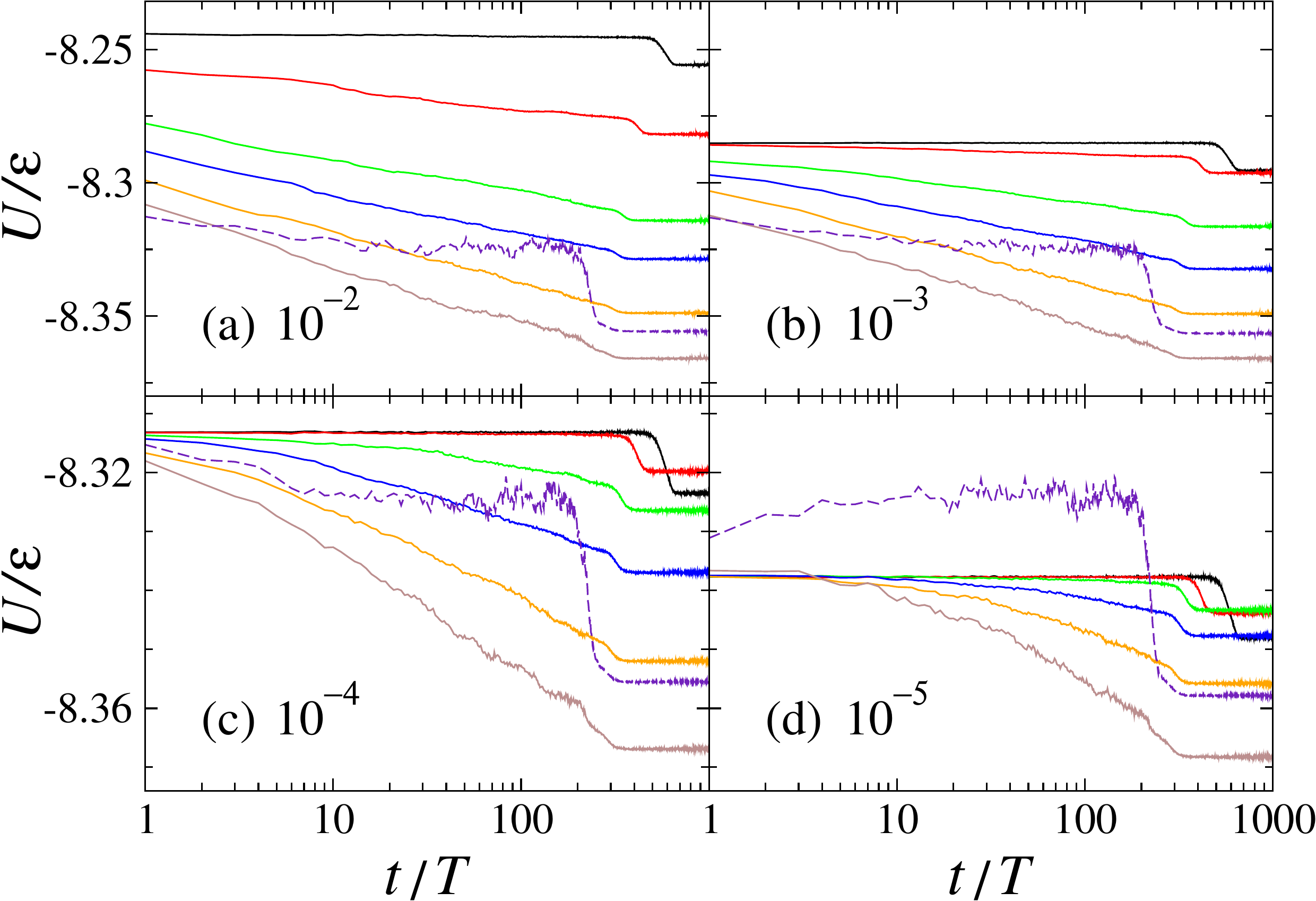}
\caption{(Color online) The potential energy minima after each
thermal cycle for glasses initially annealed with cooling rates (a)
$10^{-2}\varepsilon/k_{B}\tau$, (b) $10^{-3}\varepsilon/k_{B}\tau$,
(c) $10^{-4}\varepsilon/k_{B}\tau$, and (d)
$10^{-5}\varepsilon/k_{B}\tau$.    The potential energy for samples
at constant temperature of $0.01\,\varepsilon/k_B$ are indicated by
solid black curves. The energy minima for thermally cycled glasses
with the maximum temperature of $0.1\,\varepsilon/k_B$ (red),
$0.2\,\varepsilon/k_B$ (green), $0.25\,\varepsilon/k_B$ (blue),
$0.3\,\varepsilon/k_B$ (orange), $0.35\,\varepsilon/k_B$ (brown),
and $0.4\,\varepsilon/k_B$ (dashed indigo). Note that the vertical
scales in the upper and lower panels are different. }
\label{fig:sum_poten_min}
\end{figure}

The summary of the data for the potential energy minima after each
cycle is presented in Fig.\,\ref{fig:sum_poten_min} for binary
glasses cooled with the rates $10^{-2}\varepsilon/k_{B}\tau$,
$10^{-3}\varepsilon/k_{B}\tau$, $10^{-4}\varepsilon/k_{B}\tau$, and
$10^{-5}\varepsilon/k_{B}\tau$.   In all panels, the potential
energy for glasses aged at constant temperature of
$0.01\,\varepsilon/k_B$ are indicated by black curves (the same data
as in Fig.\,\ref{fig:poten_dens_T0.01}). Upon inspection of the data
in Fig.\,\ref{fig:sum_poten_min}, the common trend emerges for all
samples cycled with the maximum temperatures $0.1\,\varepsilon/k_B
\leqslant T_{LJ} \leqslant 0.35\,\varepsilon/k_B$; namely, the
potential energy gradually decreases during several hundred cycles
until a steady state is reached.   Typically, the onset of the
steady state with a well defined level of the potential energy
occurs sooner for higher thermal amplitudes.   Except for the cases
$T_{LJ}=0.1\,\varepsilon/k_B$ and $0.2\,\varepsilon/k_B$ in
Fig.\,\ref{fig:sum_poten_min}\,(d), the potential energy in the
steady state is lower for higher thermal amplitudes.

\vskip 0.05in


Overall, the results in Fig.\,\ref{fig:sum_poten_min} resemble the
behavior of the potential energy in glasses during cyclic shear,
where it was shown that a certain number of training cycles are
required to reach a steady state of
deformation~\cite{Sastry13,Reichhardt13,Sastry17}. In particular, it
was demonstrated that with increasing strain amplitude below yield,
the number of transient cycles increases and the potential energy
level becomes deeper, leading to steady state where particle
trajectories become exactly reversible at zero
temperature~\cite{Sastry13,Reichhardt13,Sastry17}.   However, this
analogy does not cover all aspects of thermally cycled glasses;
since MD simulations are performed at finite temperature and the
number of thermal cycles required to reach a steady state decreases
with increasing thermal amplitude, as shown in
Fig.\,\ref{fig:sum_poten_min}.   We finally note that the potential
energy curves for all glasses thermally cycled with the amplitude
$T_{LJ}=0.4\,\varepsilon/k_B$ are nearly the same after about 10
cycles (see Fig.\,\ref{fig:sum_poten_min}).

\vskip 0.05in


Additional details of the structural relaxation process during
thermal treatment can be obtained by examining the so-called
nonaffine displacements of atoms. Recall, that the nonaffine measure
quantifies a displacement of a particular atom with respect to its
neighbors during some time interval. This quantity can be evaluated
numerically as follows. First, a transformation matrix
$\mathbf{J}_i$, which describes a linear transformation of neighbors
of the $i$-th atom  during the time interval $\Delta t$, needs to be
computed~\cite{Falk98}. Then, the nonaffine measure is determined as
follows:
\begin{equation}
D^2(t, \Delta t)=\frac{1}{N_i}\sum_{j=1}^{N_i}\Big\{
\mathbf{r}_{j}(t+\Delta t)-\mathbf{r}_{i}(t+\Delta t)-\mathbf{J}_i
\big[ \mathbf{r}_{j}(t) - \mathbf{r}_{i}(t)    \big] \Big\}^2,
\label{Eq:D2min}
\end{equation}
where the sum is carried over neighboring atoms that are located
closer than $1.5\,\sigma$ to the $i$-th atom.   In the last few
years, the numerical analysis of nonaffine displacements was
performed for cyclicly
sheared~\cite{Priezjev16,Priezjev16a,Priezjev17,Priezjev18,Priezjev18a},
compressed~\cite{NVP18strload}, and thermally
cycled~\cite{Priez18tcyc} glasses.  Interestingly, it was found that
the yielding transition in poorly~\cite{Priezjev18a} and
well~\cite{Priezjev17} annealed glasses occurs after a certain
number of cycles and it is associated with the formation of a system
spanning shear band that consists of atoms with relatively large
nonaffine displacements.  On the other hand, a reduction in size of
clusters of atoms with large nonaffine displacements upon mechanical
annealing signifies an approach to a regime of reversible
deformation where most atoms return to their cages after each
cycle~\cite{Priezjev16,Priezjev16a,Priezjev17,Priezjev18,Priezjev18a,NVP18strload}.

\vskip 0.05in

\begin{figure}[t]
\includegraphics[width=12.cm,angle=0]{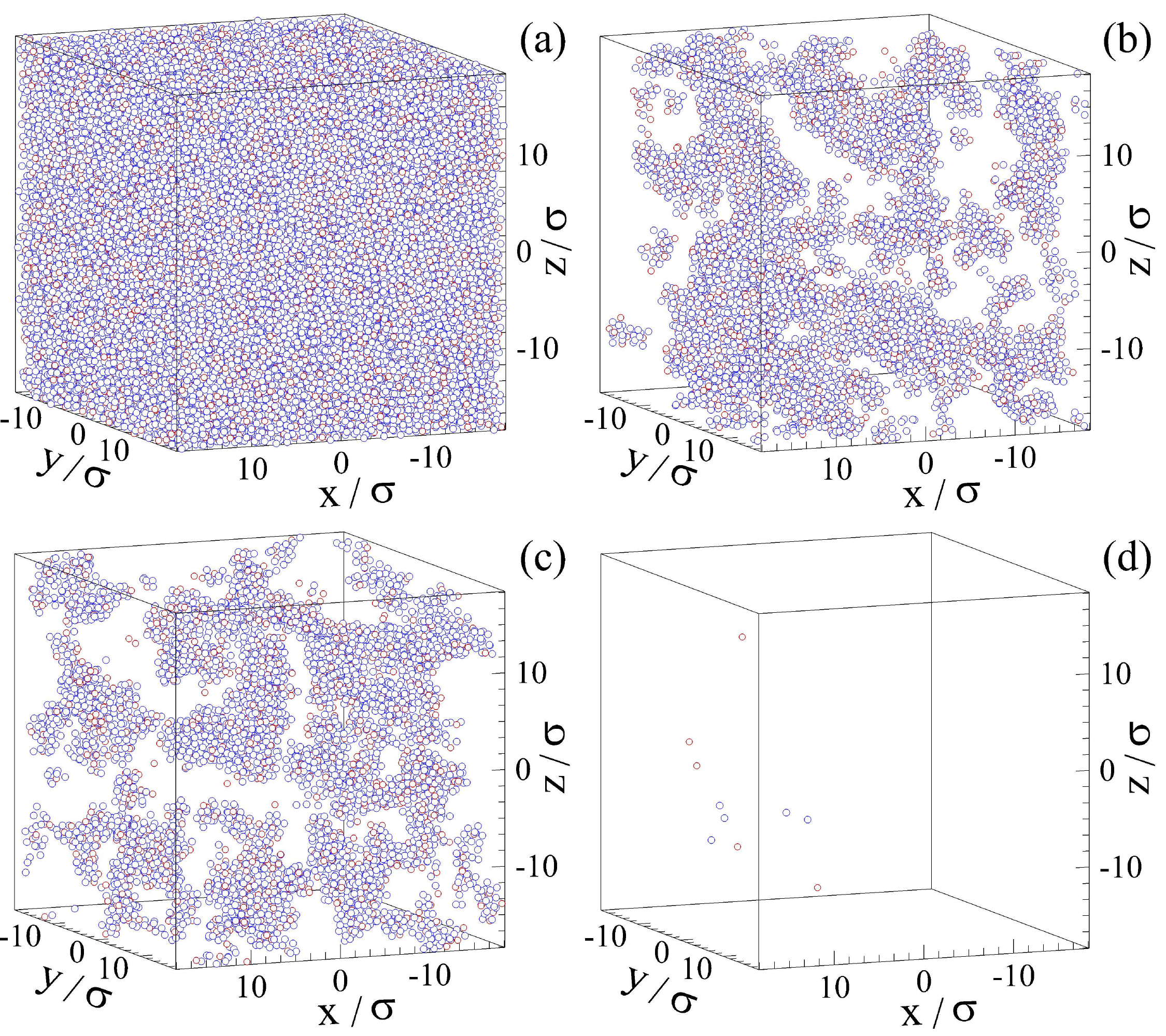}
\caption{(Color online) Spatial configurations of atoms with large
nonaffine measure (a) $D^2(0,T)>0.04\,\sigma^2$, (b)
$D^2(99T,T)>0.04\,\sigma^2$, (c) $D^2(199T,T)>0.04\,\sigma^2$, and
(d) $D^2(999T,T)>0.04\,\sigma^2$. The sample was initially annealed
with the cooling rate of $10^{-2}\varepsilon/k_{B}\tau$ and then
subjected to thermal cycling with the maximum temperature of
$0.35\,\varepsilon/k_B$.}
\label{fig:snapshot_clusters_Tm035_r10m2}
\end{figure}
\begin{figure}[t]
\includegraphics[width=12.cm,angle=0]{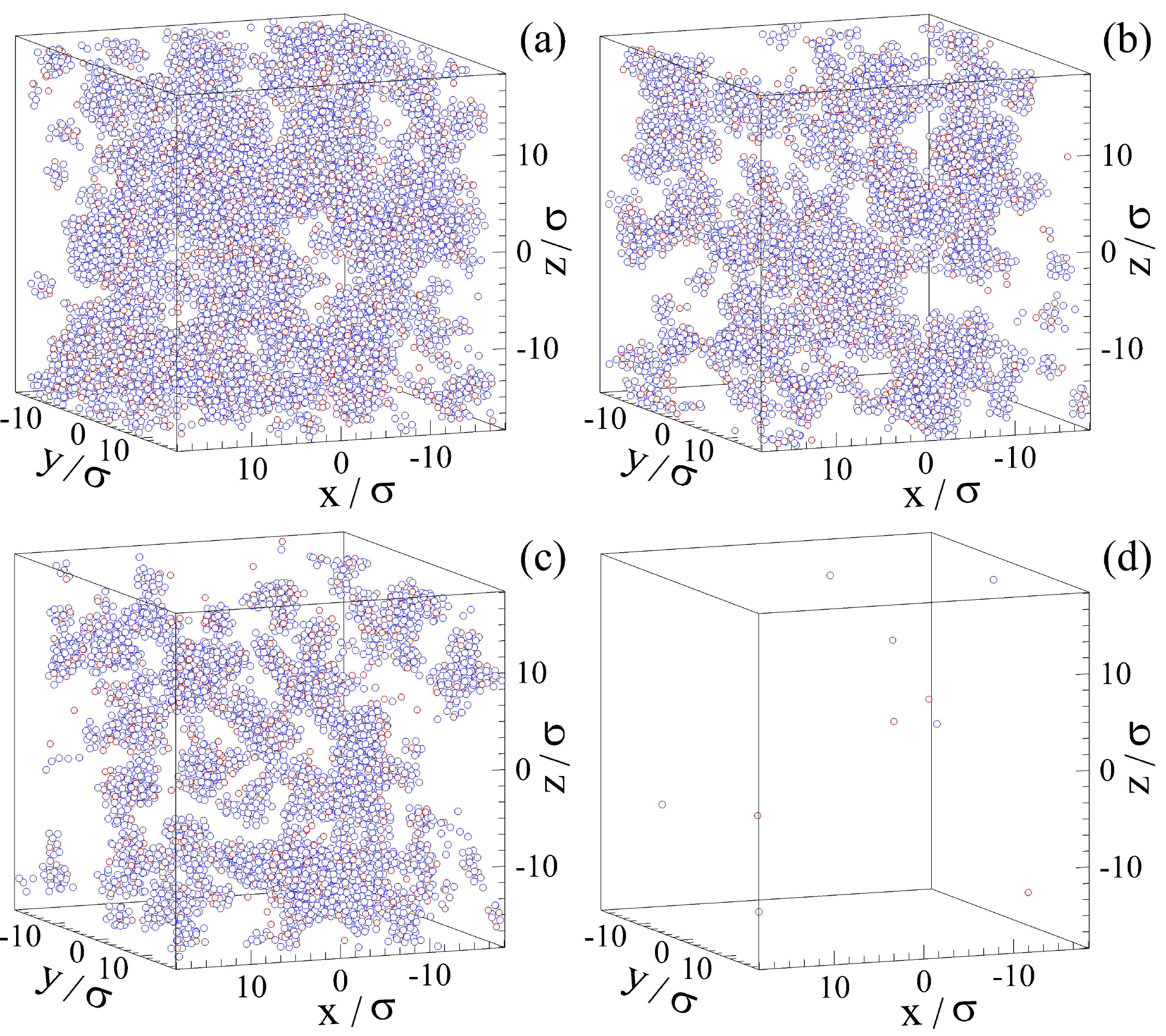}
\caption{(Color online) Positions of atoms with large nonaffile
displacements (a) $D^2(0,T)>0.04\,\sigma^2$, (b)
$D^2(99T,T)>0.04\,\sigma^2$, (c) $D^2(199T,T)>0.04\,\sigma^2$, and
(d) $D^2(999T,T)>0.04\,\sigma^2$.  The maximum temperature of
thermal oscillations is $0.35\,\varepsilon/k_B$.  The binary glass
was initially prepared with the cooling rate
$10^{-5}\varepsilon/k_{B}\tau$. }
\label{fig:snapshot_clusters_Tm035_r10m5}
\end{figure}

The sequence of spatial configurations of atoms with large nonaffine
displacements during one cycle is presented in
Fig.\,\ref{fig:snapshot_clusters_Tm035_r10m2} for poorly annealed
($10^{-2}\varepsilon/k_{B}\tau$) and in
Fig.\,\ref{fig:snapshot_clusters_Tm035_r10m5} for well annealed
($10^{-5}\varepsilon/k_{B}\tau$) glasses. In both cases, the maximum
temperature during each cycle is $0.35\,\varepsilon/k_B$.  It can be
observed in Fig.\,\ref{fig:snapshot_clusters_Tm035_r10m2} that
during the first cycle most of the atoms in the rapidly quenched
glass undergo large nonaffine displacements and leave their cages.
The typical cage size is about $r_c\approx0.1\,\sigma$. With
increasing cycle number, the characteristic size of clusters of
mobile atoms is reduced, and only a few isolated atoms are present
after the last cycle, indicating a nearly reversible dynamics.  It
was recently shown that the appearance of several mobile atoms in
glasses subjected to multiple subyield shear cycles does not
necessarily imply diffusive motion, since these atoms can jump back
and forth between neighboring cages~\cite{Priezjev16,Priezjev16a}. A
qualitatively similar behavior is reported in
Fig.\,\ref{fig:snapshot_clusters_Tm035_r10m5} for the well annealed
glass, except that the typical size of clusters of atoms with large
nonaffine displacements after the first cycle is significantly
reduced.

\vskip 0.05in


Altogether, the analysis of nonaffine displacements suggests that
during the first few hundred cycles, the thermal treatment induces
collective, irreversible rearrangements of atoms during each cycle
that leads to progressively lower potential energies.  Once a steady
state with a certain energy level is attained, the particle dynamics
becomes nearly reversible. In other words, after a certain number of
transient cycles, the glass is represented by a particular
configuration of atoms that expand upon heating but return to their
cages after each cycle. This situation is similar to the so-called
limit cycles reported in athermal quasistatic simulations of
amorphous solids, where the particle trajectories become exactly
reversible after one or more
cycles~\cite{Sastry13,Reichhardt13,Sastry17}.

\vskip 0.05in

\begin{figure}[t]
\includegraphics[width=12.0cm,angle=0]{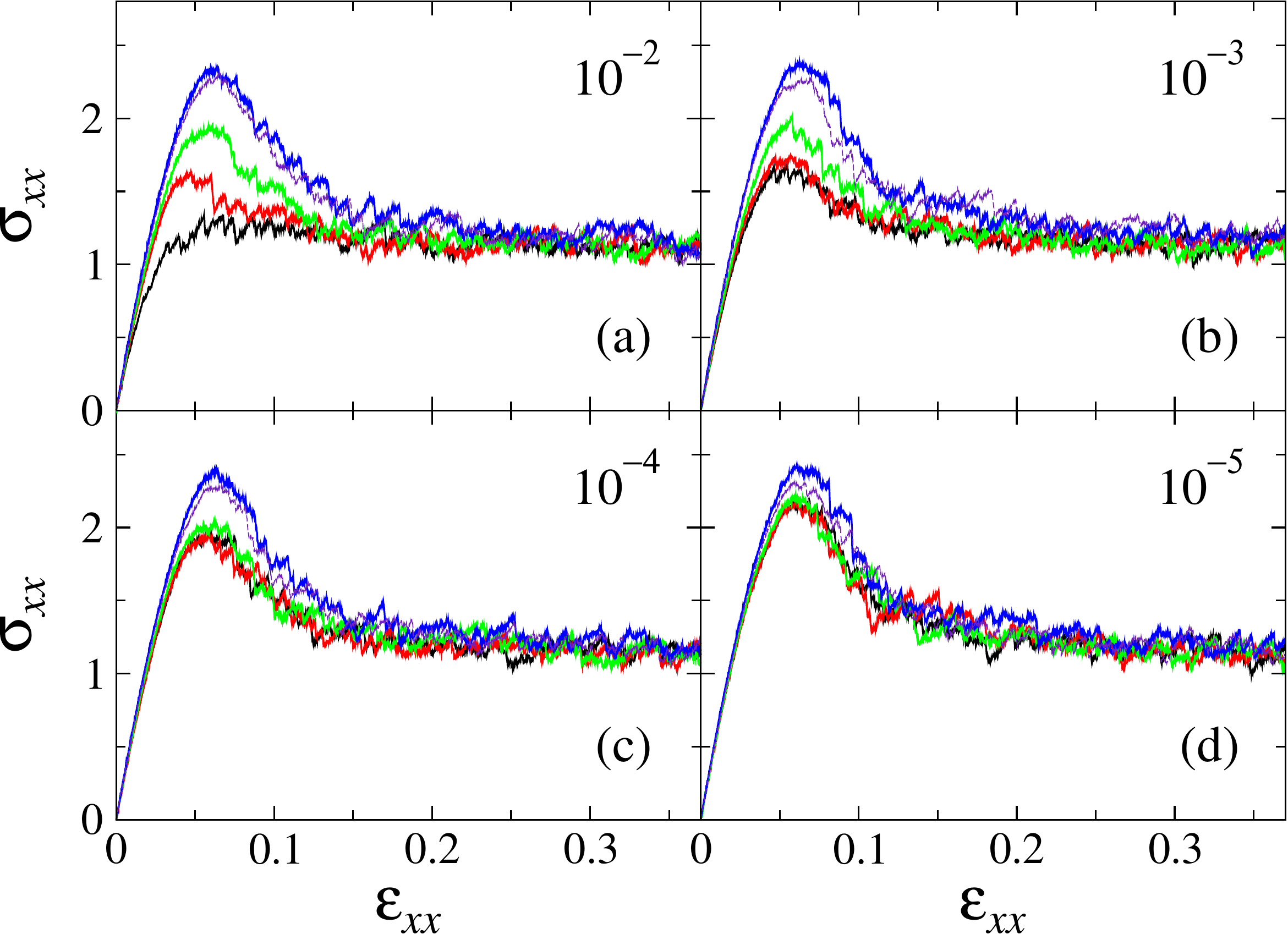}
\caption{(Color online) The tensile stress $\sigma_{xx}$ (in units
of $\varepsilon\sigma^{-3}$) versus strain for glasses prepared with
cooling rates (a) $10^{-2}\varepsilon/k_{B}\tau$, (b)
$10^{-3}\varepsilon/k_{B}\tau$, (c) $10^{-4}\varepsilon/k_{B}\tau$,
and (d) $10^{-5}\varepsilon/k_{B}\tau$.  The data for glasses aged
at $T_{LJ}=0.01\,\varepsilon/k_B$ is indicated by solid black
curves. The stress-strain response for glasses thermally cycled with
the maximum temperature $0.1\,\varepsilon/k_B$ (red),
$0.2\,\varepsilon/k_B$ (green), $0.35\,\varepsilon/k_B$ (blue), and
$0.4\,\varepsilon/k_B$ (dashed indigo). The samples were strained
with the rate $\dot{\varepsilon}_{xx}=10^{-5}\,\tau^{-1}$ at
$T_{LJ}=0.01\,\varepsilon/k_B$ and $P=0$.}
\label{fig:stress_strain}
\end{figure}

We next examine the effect of thermal cycling on mechanical
properties of binary glasses by applying a tensile strain with the
rate of $\dot{\varepsilon}_{xx}=10^{-5}\,\tau^{-1}$ at
$T_{LJ}=0.01\,\varepsilon/k_B$ and $P=0$. After the time interval
$5\times10^6\tau=1000\,T$, all thermally cycled and aged samples
were strained along the $\hat{x}$ direction up to
$\varepsilon_{xx}=0.5$. The stress-strain response is summarized in
Fig.\,\ref{fig:stress_strain} for four cooling rates and selected
values of the thermal amplitude. It can be seen that glasses that
are more slowly annealed and then aged at
$T_{LJ}=0.01\,\varepsilon/k_B$ exhibit a more pronounced yield
stress (black curves in Fig.\,\ref{fig:stress_strain}). With
increasing thermal amplitude up to $T_{LJ}=0.35\,\varepsilon/k_B$,
the yielding peak increases in all four cases shown in
Fig.\,\ref{fig:stress_strain}. Note also the the maximum stress is
approximately the same for all cooling rates when
$T_{LJ}=0.35\,\varepsilon/k_B$. Similar conclusions were obtained in
the previous study, despite that simulations were performed for
fewer values of the thermal amplitude and the steady state was not
reached~\cite{Priez18tcyc}.

\vskip 0.05in

\begin{figure}[t]
\includegraphics[width=12.cm,angle=0]{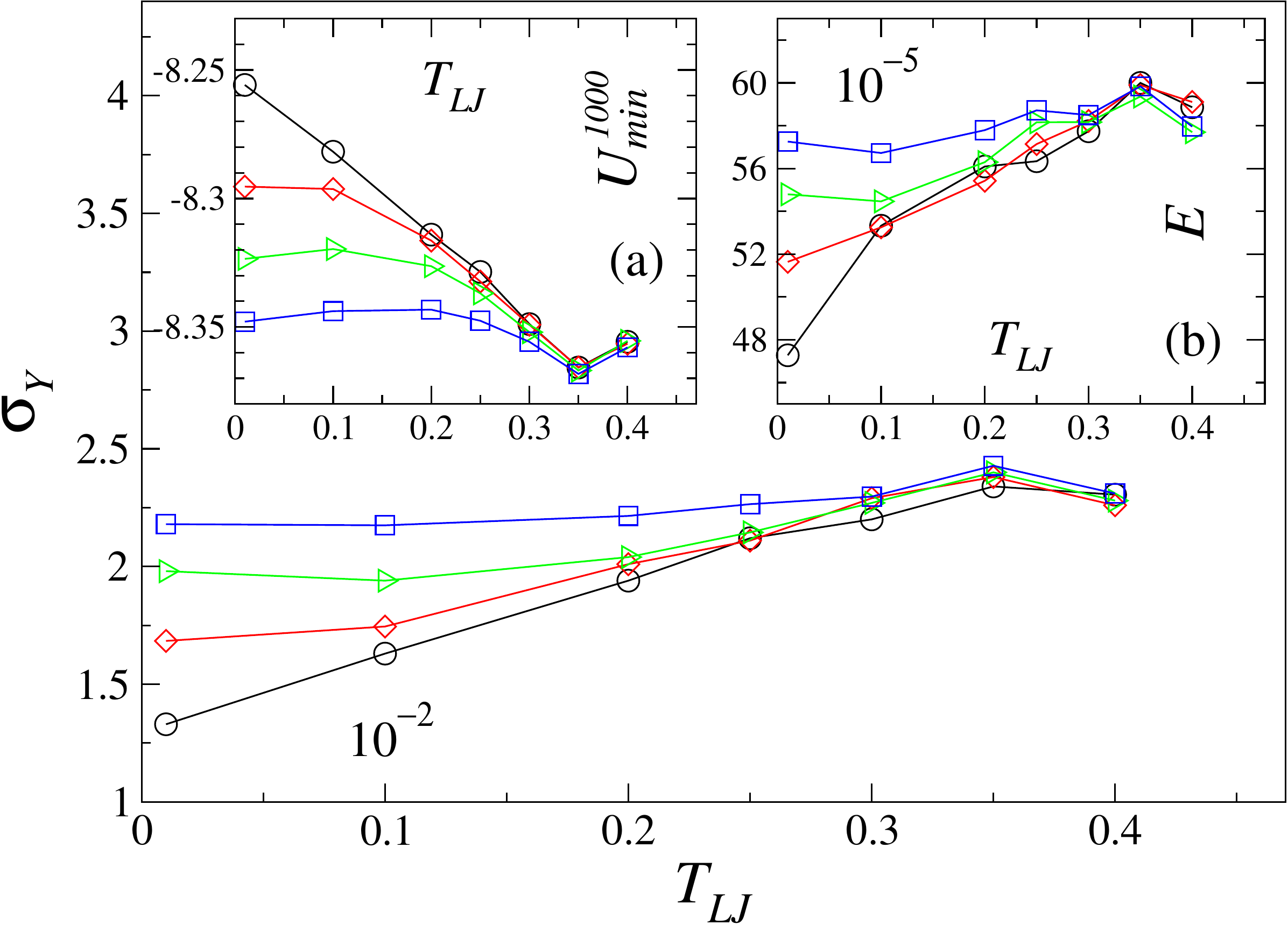}
\caption{(Color online) The yielding peak $\sigma_Y$ (in units of
$\varepsilon\sigma^{-3}$) as a function of the thermal amplitude for
cooling rates $10^{-2}\varepsilon/k_{B}\tau$ (black),
$10^{-3}\varepsilon/k_{B}\tau$ (red), $10^{-4}\varepsilon/k_{B}\tau$
(green), and $10^{-5}\varepsilon/k_{B}\tau$ (blue).  The inset (a)
shows the potential energy minima after 1000 thermal cycles.  The
inset (b) displays the elastic modulus $E$ (in units of
$\varepsilon\sigma^{-3}$) versus the thermal amplitude for the same
samples and cooling rates. }
\label{fig:yield_stress_E}
\end{figure}

We finally plot the values of the yielding peak, elastic modulus,
and the minima of the potential energy after 1000 cycles in
Fig.\,\ref{fig:yield_stress_E} as a function of the thermal
amplitude. It can be observed that with increasing thermal
amplitude, both $\sigma_Y$ and $E$ increase, except for the case
$T_{LJ}=0.4\,\varepsilon/k_B$.  Thus, the maximum of the yield
stress and elastic modulus is acquired in glasses thermally cycled
with the amplitude $T_{LJ}=0.35\,\varepsilon/k_B$.   These trends
are inversely correlated with the variation of the potential energy
minima shown in the inset (a) in Fig.\,\ref{fig:yield_stress_E}. The
origin of a slight upward curvature in $U(1000\,T)$ for the well
annealed glass at small $T_{LJ}$ is at present not clear. Given that
the potential energy in all samples in Fig.\,\ref{fig:sum_poten_min}
saturates to a constant level after several hundred cycles, we
conclude that the results for mechanical properties presented in
Fig.\,\ref{fig:yield_stress_E} should be independent of the cycle
number.

\section{Conclusions}

In summary, the effect of periodic thermal treatment on structural
relaxation and mechanical properties of binary glasses was
investigated using molecular dynamics simulations. We considered the
Kob-Andersen binary mixture annealed across the glass transition
with different cooling rates and subjected to repeated heating and
cooling cycles at constant pressure.  It was found that during a few
hundred transient cycles, the potential energy minima gradually
decrease toward a steady state with the lowest energy level, which
depends on the thermal amplitude.  Moreover, with increasing thermal
amplitude up to a certain value, the number of transient cycles
decreases and the level of the potential energy is reduced. The
structural relaxation proceeds via collective, nonaffine
displacements of atoms that are organized into clusters. In
contrast, the particle dynamics in steady state becomes nearly
reversible and the system returns to the same energy state after
each cycle. The results of tensile tests after one thousand cycles
revealed that the maximum values of the elastic modulus and yielding
peak occur at the same thermal amplitude regardless of the annealing
history.

\vskip 0.05in

In the future, it might be instructive to investigate whether the
details of thermal treatment, such as the reference temperature,
oscillation period, system size, and choice of atomistic model, will
affect the main conclusions regarding the potential energy states
and mechanical properties of amorphous materials.  One of the
findings of the present study is the appearance of well-defined
steady states with particular energy levels and nearly reversible
dynamics.  This, in turn, opens the question of memory formation in
such systems similar to memory effects reported in periodically
driven glasses~\cite{Keim18}.  For example, thermally cycled glasses
with a particular thermal amplitude might continue to expand and
contract reversibly if the thermal amplitude is reduced or instead
undergo structural relaxation if the thermal amplitude is increased.

\section*{Acknowledgments}

Financial support from the National Science Foundation (CNS-1531923)
is gratefully acknowledged. The article was prepared within the
framework of the Basic Research Program at the National Research
University Higher School of Economics (HSE) and supported within the
framework of a subsidy by the Russian Academic Excellence Project
`5-100'. The molecular dynamics simulations were performed using the
LAMMPS numerical code developed at Sandia National
Laboratories~\cite{Lammps}. Computational work in support of this
research was performed at Wright State University's Computing
Facility and the Ohio Supercomputer Center.


%

%

%

%
%

%
%

%
%

%

%

%

%

\bibliographystyle{prsty}

\end{document}